\documentclass[prl,amsmath,amssymb,twocolumn]{revtex4}

\usepackage{graphicx}
\usepackage{subfigure}
\usepackage{adjustbox}
\usepackage{bm}
\usepackage{color}
\usepackage{braket}
\usepackage{standalone}
\usepackage{multirow}
\usepackage{tikz}
\usepackage{mathrsfs}
\usepackage{dsfont}
\usepackage[colorlinks,bookmarks=true,citecolor=blue,linkcolor=red,urlcolor=blue]{hyperref}

\newcommand{\bea}{\begin{eqnarray}}
\newcommand{\eea}{\end{eqnarray}}

\begin{document}

\title{Non-Hermitian Weyl Physics in Topological Insulator Ferromagnet Junctions}

\author{Emil J. Bergholtz$^1$ and Jan Carl Budich$^2$}

\affiliation{$^1$Department of Physics, Stockholm University, AlbaNova University Center, 106 91 Stockholm, Sweden\\
$^2$Institute of Theoretical Physics, Technische Universit\"{a}t Dresden, 01062 Dresden, Germany
}
\date{\today}

\begin{abstract}
We introduce and investigate material junctions as a generic and tuneable electronic platform for the realization of exotic non-Hermitian (NH) topological states of matter, where the NH character is induced by the surface self-energy of a thermal reservoir. As a conceptually rich and immediately experimentally realizable example, we consider a three-dimensional topological insulator (TI) coupled to a ferromagnetic lead. Remarkably, the symmetry protected TI is promoted in a dissipative fashion to a non-symmetry protected NH Weyl phase with no direct Hermitian counterpart and which exhibits robustness against any perturbation. The transition between a gapped phase and the NH Weyl phase may be readily tuned experimentally with the magnetization direction of the ferromagnetic lead. Given the robustness of this exotic nodal phase, our general analysis also applies to, e.g., a two-dimensional electron gas close to criticality in proximity to a ferromagnetic lead. There, the predicted bulk Fermi arcs are directly amenable to surface spectroscopy methods such as angle-resolved photoemission spectroscopy.    
\end{abstract}

\maketitle

{\it Introduction.---}
While Hermiticity is a basic requirement on the Hamiltonian governing the dynamics of an isolated quantum system, non-Hermitian (NH) effective Hamiltonians have become a ubiquitous tool with applications ranging from dissipative classical optical and mechanical systems to various open quantum systems \cite{bender,ganainymakriskhajavikhanmusslimanirotterchristodoulides,lujoannopoulossoljacic,rotter,BerryDeg,Heiss}. Recently, the interest in NH Hamiltonians has been further fueled by conceptual insights relating to exceptional degeneracies that lead to novel (topological) phases of matter with no direct counterpart in the conventional Hermitian realm \cite{NHarc,koziifu,carlstroembergholtz,shenzhenfu,asymhop2,holler,kawabata19}. In particular, a new system of symmetry-protected NH topological phases extending the celebrated periodic table of topological insulators known from Hermitian realm has been identified 
\cite{gong,jan,kawabata,zhou,Okugawa,schomerus,yuce,leykambliokhhuangchongnori,lieu2,MaAlVaVaBeFoTo2018,Das}, and fundamental amendments to the occurrence of topologically protected surface states have been reported 
\cite{KuEdBuBe2018,lee,Xi2018,yaowang,leethomale,KuDw2018,borgnia}. So far, the main platforms for the observation of these intriguing phenomena have been photonic systems subject to gain and loss \cite{wiemannkremerplotniklumernoltemakrissegevrechtsmanszameit,NHtransition,EPringExp,NHexp2,NHlaser,asymhop1}, even though NH self-energies in electronic systems are in principle known to be capable of inducing genuinely NH phenomena as well \cite{koziifu,yoshidapeterskawakmi,Zyuzin2018,papaj,Yoshida,Zyuzin2019,Moors,MadridMajoranaEP,MajoranaEPPreprint,molina,Philip,Pikulin}.

\begin{figure}[t]
	\includegraphics[width=\linewidth]{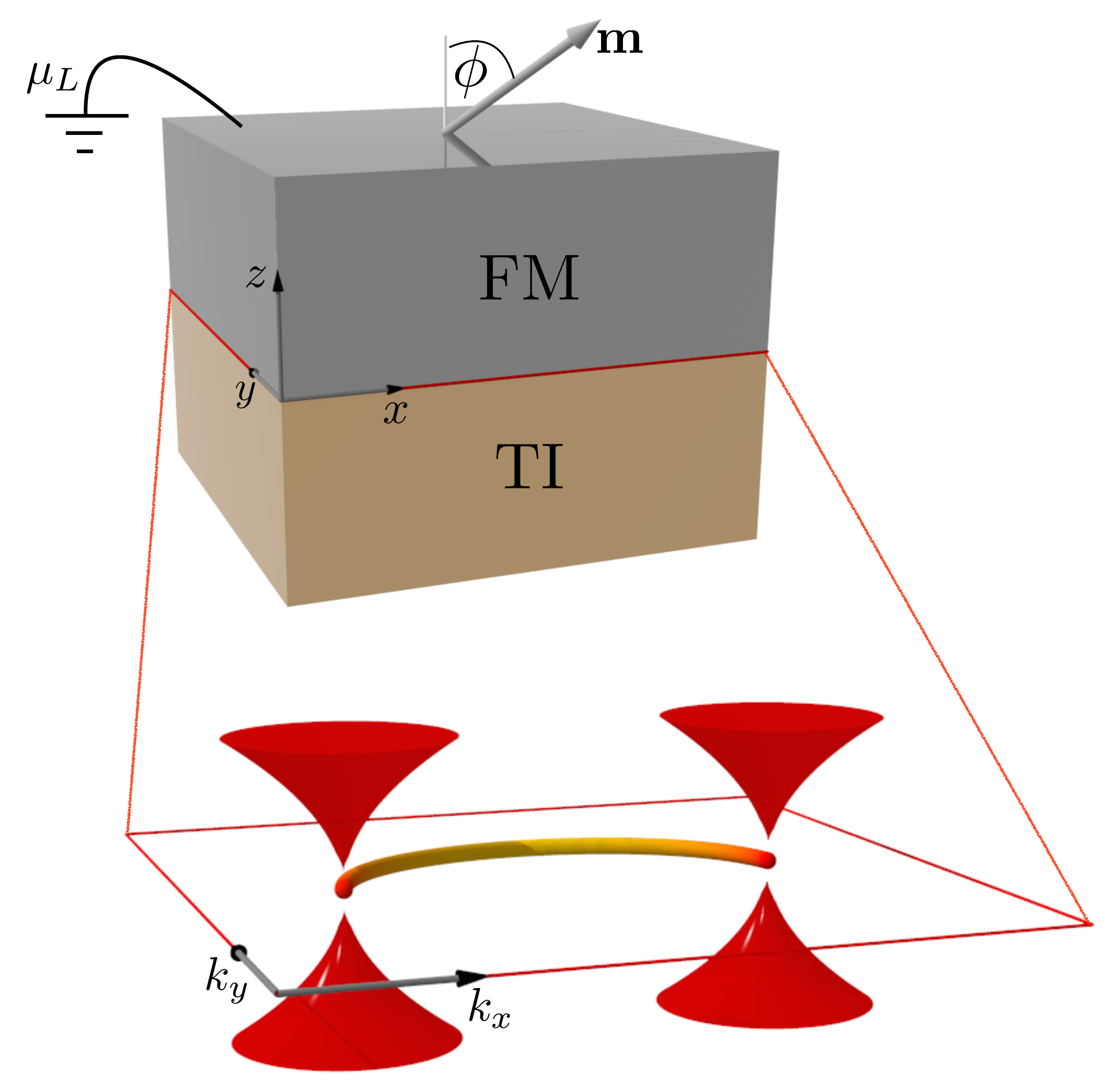}
	\caption{Top: Schematic of a 3D topological insulator (TI) coupled to a ferromagnetic metallic lead (FM) with magnetization $\mathbf m$, and in equilibrium at chemical potential $\mu_L$. The polar angle of the magnetization direction is denoted by $\phi$, and the red frame marks the interface (material junction). Bottom: Illustration of the interface between the two materials. Upon tuning $\phi$, in the NH effective Hamiltonian $H_{\textrm{NH}}$  (see Eq.~(\ref{eqn:generalHNH})) describing the junction, a pair of exceptional points with a characteristic square root dispersion appears, that are connected by an open Fermi surface.}
	\label{fig1}
\end{figure}

In this work, we propose and study material junctions as a simple and generic {\emph{electronic}} setting for realizing NH topological phases. There, one side of the junction is considered to be a thermal reservoir (lead) which induces a self-energy on the surface of the system thus leading to the effective NH system Hamiltonian
\begin{align}
H_{\textrm{NH}}=H+\Sigma_L^r(\omega=0)
\label{eqn:generalHNH}
\end{align}
where $H$ is the Hermitian Hamiltonian of the isolated system and $\Sigma_{\textrm{L}}^r(\omega=0)$ denotes the retarded self-energy at the chemical potential, reflecting the coupling to the lead. While the self-energy in general is frequency dependent, approximating it with its value right at the Fermi energy is justified for our present analysis of nodal points \cite{Foot1}. As a concrete immediately experimentally feasible setup, we focus first on a three-dimensional (3D) topological insulator (TI) \cite{hasankane,qizhang}
 coupled to a metallic ferromagnetic (FM) lead (see Fig.~\ref{fig1} for an illustration) \cite{Philip}. We show that this system exhibits a topological phase transition that is naturally controlled by the magnetization direction $\mathbf m$ of the FM lead. In particular, at a critical polar angle $\phi_c$ of $\mathbf m$, a NH Weyl phase occurs, featuring two separated exceptional points that are connected by a Fermi arc resulting in an emergent double Riemann sheet topology.

Quite remarkably, in the proposed setup, the dissipative perturbation represented by the surface self-energy promotes the time reversal symmetry (TRS) protected semimetallic surface Dirac cone of the 3D TI to a non-symmetry protected {\emph{metallic}} NH topological phase that is stable to {\it any} perturbation. To illustrate this behavior, we show that even if the Hermitian surface Dirac cone of the 3D TI is gapped out by symmetry breaking terms, the coupling to the ferromagnetic lead overcomes these imperfections, pushing the system into the extended NH topological Weyl phase in a dissipative fashion. Since this stable phase does not require fine-tuning, it can be realized even in simpler systems, where the role of the 3D TI is played by a near-critical 2D electron gas (2DEG) [see Fig.~\ref{figtransport} (b)], thus alleviating the requirement of a symmetry protected nodal point in the unperturbed Hermitian spectrum, and making the NH Weyl phase directly amenable to surface spectroscopy.

{\textit{Microscopic model. --}}
We study a 3D TI cubic lattice model coupled to a ferromagnetic lead. The 3D TI may in reciprocal space be described by the four-band Bloch Hamiltonian \cite{FuBerg2010,qizhang}
\begin{align}
&H_{\textrm{TI}}(\mathbf k) = \left(M-\cos(k_x)-\cos(k_y)-\cos(k_z)\right)\tau_x\sigma_0\nonumber\\
&+\lambda(\sin(k_y)\tau_z\sigma_x-\sin(k_x)\tau_z\sigma_y+\sin(k_z)\tau_y\sigma_0),
\label{eqn:sysham}
\end{align}
where $\sigma$ ($\tau$) denote the standard Pauli matrices in spin (orbital) space, $\lambda$ is a spin-orbit coupling strength, $M$ is the Dirac mass parameter, length is measured in units of the lattice constant, and energy in units of the hopping strength. The FM lead is in reciprocal space described by the Bloch Hamiltonian 
\begin{align}
H_L(\mathbf k) = &-2t\left(\cos(kx)+\cos(ky)\right)\sigma_0-2t_z\cos(k_z)\sigma_0\nonumber\\
&-\mu_L\sigma_0+\mathbf m\cdot \boldsymbol\sigma, 
\label{eqn:leadham} 
\end{align}  
where $t$ is the hopping strength in the $xy$-plane, $t_z$ denotes the hopping strength in the $z$-direction, the chemical potential is represented by $\mu_L$, and $\mathbf m$ is the magnetization. We find that an anisotropic band mass in the lead ($t_z \ne t$) may quantitatively support the NH Weyl phase, but our main qualitative results remain unchanged in the simplest case $t=t_z$.  To model the coupling between the system and the lead, we consider a half-space geometry in the $z$-direction, where $z>0$ is the realm of the 3D TI, while the semi-infinite lead resides in the $z\le0$ half space. Accounting for this half-space geometry, we switch to a real space tight-binding description in the $z$-direction and interface the system with the lead by the spin-independent hopping strength $V_{\textrm{SL}}$ between the last site of the lead ($z=0$) and the first site of the 3D TI in ($z=1$). The spin dependent self-energy $\Sigma_{\textrm{L}}^r(0)=V_{\textrm{SL}} G_L^r(0) V_{\textrm{SL}}^\dag$ with the retarded lead Green's function $G_L^r$ is then readily expressed analytically as \cite{CunibertiReview}
\begin{align}
\Sigma_{\textrm{L},\sigma}^r(0)= &\frac{\lvert V_{\textrm{SL}} \rvert^2}{t_z}\left(\kappa_\sigma -\sqrt{\kappa_\sigma^2-1}[\text{sgn}(\kappa_\sigma)\theta(\lvert\kappa_\sigma\rvert-1)]  \right)\nonumber\\
& - i \frac{\lvert V_{\textrm{SL}} \rvert^2}{t_z}\sqrt{1-\kappa_\sigma^2}\theta(1-\lvert\kappa_\sigma\rvert),\quad\sigma=\pm,
\label{eqn:selfen}
\end{align}
where $\kappa_{\sigma=\pm}=\left(\mu_L-2t\left(\cos(k_x)+\cos(k_y)\right)\pm \lvert \mathbf m\rvert\right)/(2t_z)$, and $\sigma=\pm$ labels the spin-eigenstates along the quantization axis $\mathbf m/\lvert \mathbf m\rvert$ of the FM lead.

\begin{figure}[t]
	\includegraphics[width=\linewidth]{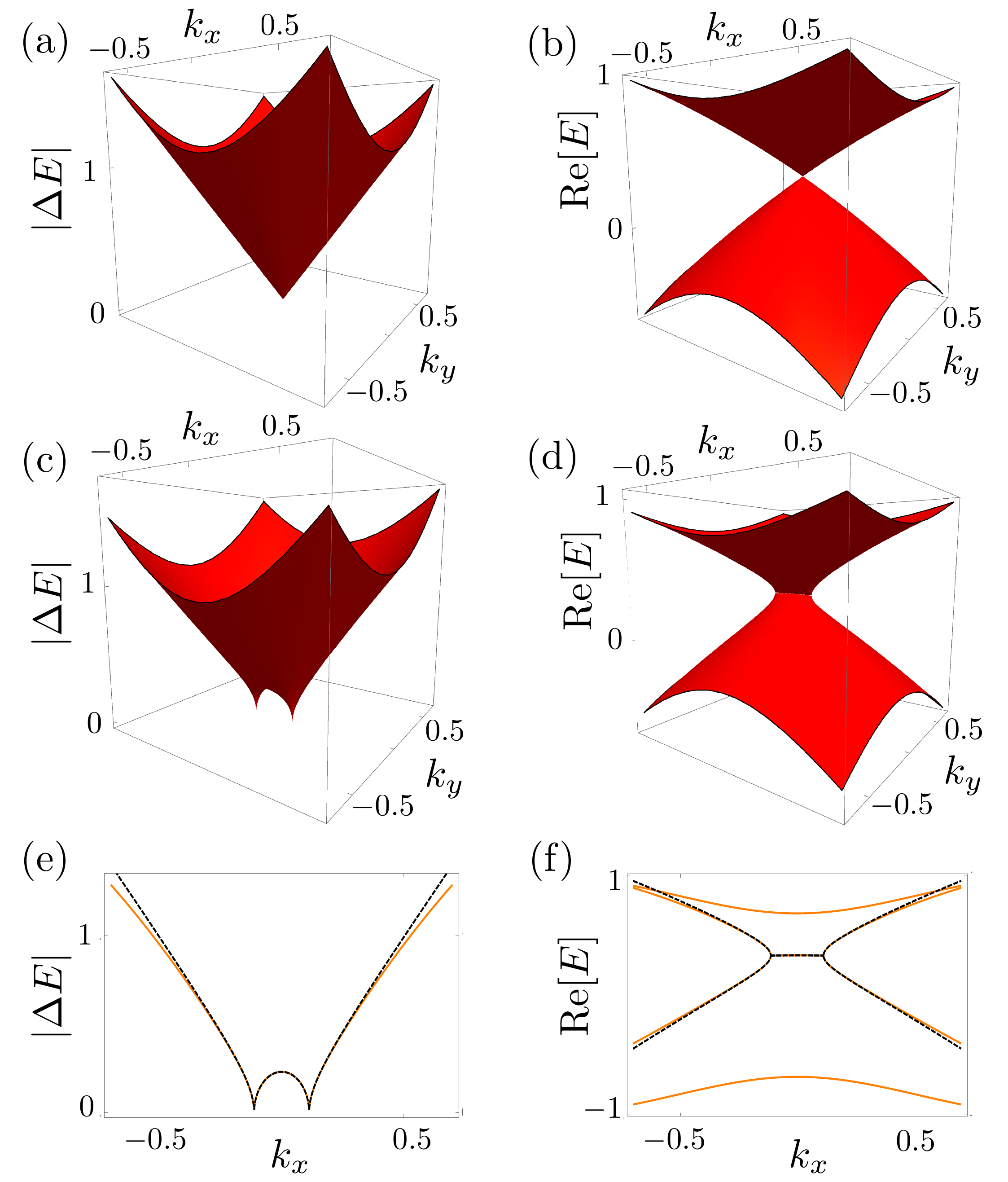}
	\caption{Panels (a)-(d): Spectra of the effective two-band model (\ref{eqn:heff}). Panels (e)-(f):  Quantitative comparison with the full microscopic model [see Eqs. (\ref{eqn:generalHNH}-\ref{eqn:selfen})]. The spectrum in (a) and (b) corresponds to the case without magnetization, i.e. $|\mathbf m|=0$, leading to a conventional Dirac dispersion. In (c)-(f) there is a finite magnetization $|\mathbf m|=0.5$ making an angle $\phi=\pi/2$ relative to the $z$-axis ($\mathbf{m}\sim \hat e_x$) which leads to a NH Weyl phase with exceptional points with a characteristic square root dispersion in the absolute value of the gap ((c) and (e)) which are joined by a Fermi arc node in the real part of the spectrum (see (d) and (f)). In (e) and (f) the black dashed (solid orange) lines indicate the effective (full microscopic) model, respectively.  For these examples we have used $t=1, t_z=1.3, \mu_L=2$ and $V_{SL}=t_z/\sqrt{3}$. In (e) and (f), the value of $k_y=-0.083$ is fixed so as to obtain a 1D cut through in the BZ that contains the exceptional points and the characteristic Fermi arc.}
	\label{fig2}
\end{figure}

{\it Effective model.---}
While we carefully verify our predictions by comparison to the full lattice model [see Eqs. (\ref{eqn:generalHNH}-\ref{eqn:selfen}) and Fig.~\ref{fig2}(e) and \ref{fig2}(f)], we find that all qualitative physical properties of the TI-FM interface can be understood from the effective model Hamiltonian  
\begin{eqnarray}
&\tilde H=\lambda(k_y\sigma_x\!-\!k_x\sigma_y)+\Sigma_{L}^r(0)-B \sigma_z\equiv\epsilon_0+\mathbf d\!\cdot\! \boldsymbol \sigma, 
\label{eqn:heff}
\end{eqnarray}
where $\epsilon_0\in \mathbb C$ and $\mathbf d = \mathbf d_R\!+i\mathbf d_I$ with $\mathbf d_R,\mathbf d_I \in \mathbb R^3$. Equation (\ref{eqn:heff}) may either describe the chiral surface Dirac cone of the 3D TI with a TRS breaking Hermitian term $B$, or alternatively the low-energy theory of a near-critical 2DEG with a residual mass term $B$, on which we elaborate further below. In both scenarios, $\Sigma_{L}^r(0)$ represents the full self-energy induced by the lead (see Eq.~\ref{eqn:selfen}).

The effective model (\ref{eqn:heff}) has complex energy eigenvalues    
$E_\pm=\epsilon_0\pm \sqrt{\mathbf d_R^2 - \mathbf d_I^2 +2i \mathbf d_R\cdot\mathbf d_I}$ and exhibits exceptional degeneracies at $E=\epsilon_0$ when 
\begin{equation}\mathbf d_R^2 = \mathbf d_I^2 \ \ {\rm and} \ \ \mathbf d_R\cdot\mathbf d_I=0 \label{epcond}\end{equation} 
are simultaneously satisfied. Since $\mathbf d_R$ and $\mathbf d_I$ are functions of two continuous variables, $k_x$ and $k_y$, the solutions to Eq. (\ref{epcond}) are generic, point-like, appear in pairs, and are stable much as Hermitian Weyl points are stable in three dimensions. 
At the exceptional points, however, not only the eigenvalues but also eigenvectors coalesce and the complex dispersion has a characteristic square root dispersion as depicted in Fig.~\ref{fig1}. Moreover, on the closed curve of solutions to $\mathbf d_R\cdot\mathbf d_I=0$ in reciprocal space the energy is purely real or imaginary depending on the sign of $\mathbf d_R^2 - \mathbf d_I^2$ leading to a ${\rm Re} [E]=0$ Fermi arc connecting the exceptional points.

\begin{figure}[htp]
	\includegraphics[width=\linewidth]{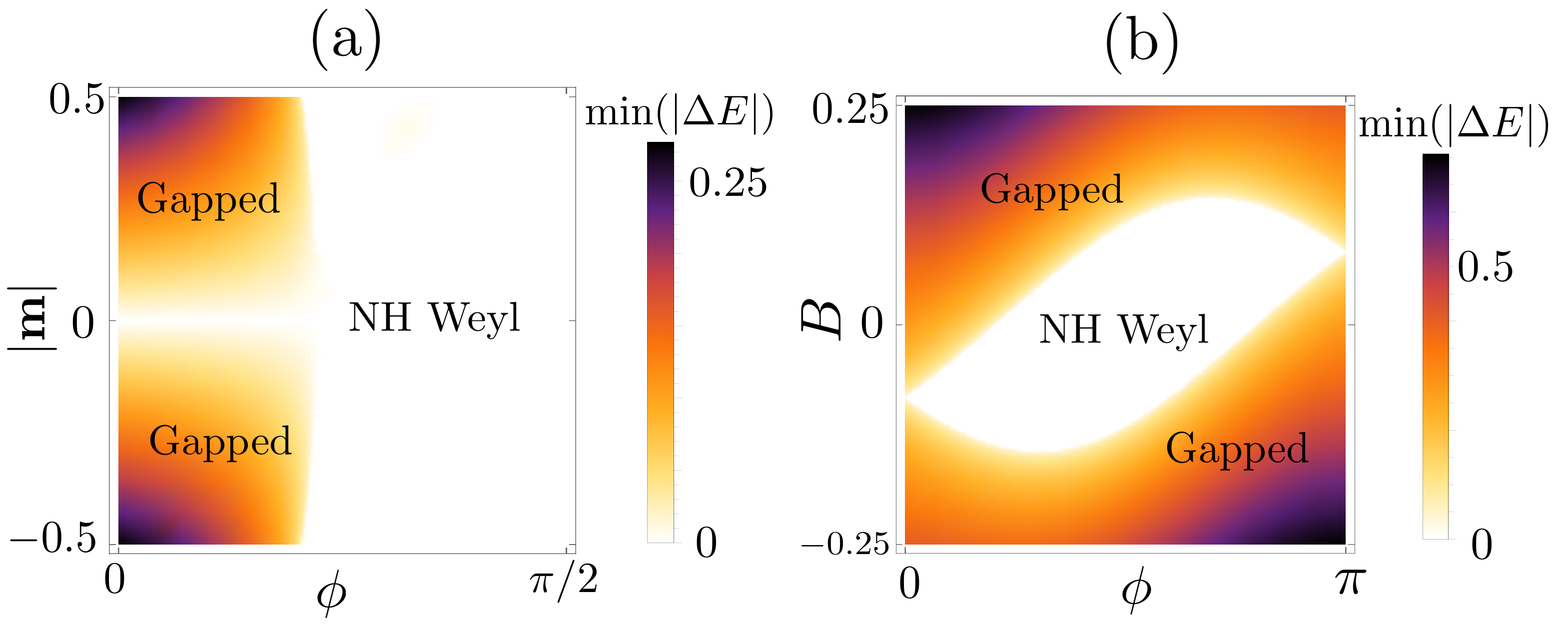}
	\caption{Phase diagram of the effective model (\ref{eqn:heff}), defined via the minimum absolute value of the gap extracted in the physically relevant region $|k_x|, |k_y|\leq 0.7$ [see Fig.~\ref{fig2}(e) and \ref{fig2}(f)]. The white zero gap region characterizes the (NH) Weyl phase, which for finite $\lvert\mathbf{m}\rvert$ emerges when the tilting angle $\phi$ of the lead magnetization exceeds a critical value $\phi_c$ as shown in (a) for $B=0$. In (b), we consider the effect of finite $B$ at fixed $|\mathbf m|=0.5$. At $\phi \ne 0,\pi$, a non-symmetry protected NH Weyl phase stable to any perturbation manifests through the extended white region. In both plots the model parameters are $t=1, t_z=1.3, \mu_L=2$ and $V_{SL}=t_z/\sqrt{3}$.}
	\label{fig3}
\end{figure}

We start with the unperturbed case of $B=0$, where we have $\mathbf d_R=\lambda(k_y,-k_x,0) + \alpha \mathbf m,~\mathbf d_I=\beta \mathbf m$, with $\alpha={\rm Re}[\Sigma_{\textrm{L},+}^r(0)-\Sigma_{\textrm{L},-}^r(0)]/2$ and $\beta={\rm Im}[\Sigma_{\textrm{L},+}^r(0)-\Sigma_{\textrm{L},-}^r(0)]/2$ [cf. Eq. (\ref{eqn:selfen})]. Physically, a finite $\alpha$ may be understood as a Hermitian TRS breaking perturbation induced by the FM lead, whereas $\beta$ is the strength of the spin-dependent NH perturbation that leads to the intriguing NH phenomena reported in this work. An intuitive picture is that the magnetic character of the lead does not just entail a simple level broadening but rather a spin-dependent matrix structure in the anti-Hermitian part of the self-energy. This leads to a sufficiently generic non-Hermitian effective Hamiltonian to create a pair of exceptional points. The first condition in Eq. (\ref{epcond}) does indeed have generic solutions when $\alpha^2\cos^2(\phi)<\beta^2$, with $\cos^2(\phi)=m_z^2/ |\mathbf m|^2$, which defines the regime of physical interest where the NH perturbation due to the lead dominates the Hermitian one. Furthermore, the second condition in Eq. (\ref{epcond}) implies that the angle $\phi$ has to be finite, i.e. exceptional points require that the magnetization is not perpendicular to the interface. Since $k_x,k_y$ are bounded in the microscopic lattice model, for a given parameter set we find that there is a critical angle $\phi_c$ that marks the onset of the non-Hermitian Weyl physics [see Fig. \ref{fig3}(a)], where the Dirac cone at $|\mathbf m|=0$ is extended to a finite gapless (white) region corresponding to the NH Weyl phase.

Now allowing for finite $B$ in Eq.~(\ref{eqn:heff}), we demonstrate how the symmetry protected surface Dirac cone of the TI is promoted to the generically stable NH Weyl phase [see Fig. \ref{fig3}(b)]. Since this nodal phase persists even in the presence of the explicitly TRS breaking perturbation $B$ that would gap out the Dirac surface theory of the Hermitian TI, a near-critical 2D Dirac system is found to be sufficient as a starting point for realizing NH Weyl physics. This suggests the aforementioned alternative interpretation of Eq.~(\ref{eqn:heff}), namely as a 2DEG close to a transition between a 2D TI (Chern insulator) and a trivial insulator. In both cases, varying the magnetization vector $\mathbf m$ in the lead, the non-Hermitian terms can overcome a small gap, thus driving the system into the topologically stable NH Weyl phase in a large region of the phase diagram.

{\it Experimental signatures.---} The salient property of the NH Weyl phase discussed in this work is its metallic character which is remarkable in two regards. First, it occurs in a symmetry breaking environment where one would intuitively expect the TI surface to become insulating. Second, the NH Weyl phase exhibits extensive surface conductance due to the finite length of the Fermi arc in reciprocal space [see Fig. \ref{fig2}], whereas a Hermitian Dirac cone represents an isolated point degeneracy and thus is {\emph{semimetallic}}. A possible geometry for probing the surface conductance of the NH Weyl phase is shown in Fig.~\ref{figtransport}(a): A 3D TI surface underneath a FM lead representing the NH Weyl region is interfaced with an uncovered 3D TI surface that is connected to a normal lead (NL). Since the FM lead generically entails a finite energy shift (see Fig. \ref{fig2}), tuning the chemical potential of the TI to the NH Fermi arcs automatically moves the Fermi energy of the free TI surface states away from the Dirac point, thus making them metallic as well. While this setting thus in principle allows for measuring a metallic conductance of the TI surface despite the TRS breaking FM lead, it is fair to say that in practice separating bulk from surface transport properties in 3D TI materials remains challenging. Hence, the most established experimental signatures of 3D TIs so far have been spectroscopic properties as observed in angle resolved photoemission spectroscopy (ARPES) experiments. However, in the proposed 3D TI-FM interface the NH Weyl phase is covered by the bulk TI from one side and the bulk FM lead from the other side (see Fig.~\ref{figtransport}(a)).

To overcome this issue and make NH Weyl physics discussed here amenable to surface spectroscopy, the stability of the NH Weyl phase against general perturbations comes to the rescue in the sense that there is no need to start from a precisely gapless Dirac cone as a Hermitian part of the Hamiltonian. Thus, the 3D TI with its symmetry protected gapless surface state, may simply be replaced by a 2DEG tuned fairly close (but not necessarily exactly) to a critical point in the form of a {\emph{bulk}} 2D Dirac cone, as realized e.g. at the transition between a Chern insulator and a trivial phase. By putting this near-critical 2DEG on top of a FM lead, quite generically a 2D NH Weyl phase that can be experimentally analyzed by surface spectroscopy methods [see Fig.~\ref{figtransport}(b)] will be stabilized in a finite parameter range [see Eq.~(\ref{eqn:heff}) and Fig.~\ref{fig3}(b)].

\begin{figure}[htp]
	\includegraphics[width=\linewidth]{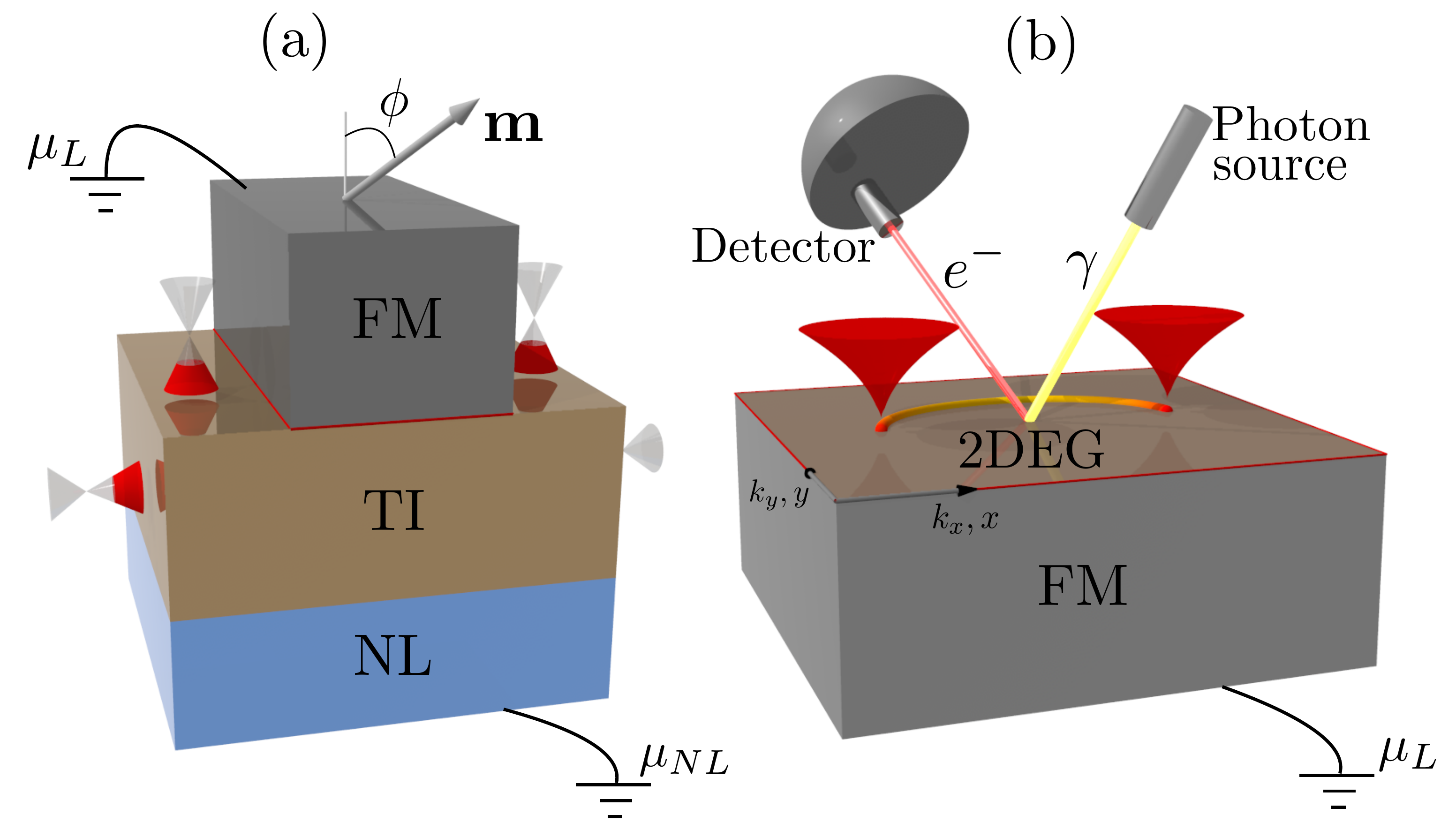}
	\caption{Schematic illustrations of experimental setups in which the NH Weyl phase (see Eq.~(\ref{eqn:heff}) and Fig.~\ref{fig3}) can be probed. (a) 3D TI in proximity to a ferromagnetic lead in which surface transport remains metallic in the NH Weyl regime. (b) Alternative setup up with a near critical two-dimensional electron gas (2DEG), where the NH Weyl phase can be identified by surface spectroscopy measurements such as ARPES.}
	\label{figtransport}
\end{figure}

{\it Discussion.---}
In this work we have put forward material junctions as a generic and tuneable platform for realizing novel non-Hermitian phases of matter. In particular, our explicit example of a three-dimensional topological insulators coupled to a ferromagnetic lead shows how opening the system to a thermal reservoir can promote a symmetry protected Hermitian topological phase into a more robust NH state of matter that does not rely on any symmetry, and which also does not have a direct analog in the Hermitian realm. 

In this context, we would like to contrast the NH Weyl phase realized in our proposed setup with conventional Hermitian Weyl semimetals. While the Hermitian Weyl systems feature discrete band touching points, the dispersion is always analytic and the Hamiltonian stays diagonalizable at the degenerate points. In contrast, the NH dispersion is non-analytic and at the exceptional points two eigenvectors coalesce thus rendering the Hamiltonian non-diagonalizable. Moreover, the corresponding NH Fermi arcs are different from the surface Fermi arcs of Weyl semimetals in that they constitute open bulk Fermi surfaces that are only augmented to closed curves by a complementary ${\rm Im} [E]=0$ arc.

In photonic systems, parity-time (PT) symmetry, which is realized by a balancing gain and loss in a classical optical setup, plays an important role. This is because PT symmetry leads to an extended regime of real energy eigenvalues, known as the PT-unbroken phase. In our present electronic context, causality of the system-lead interaction restricts the NH terms in the effective Hamiltonian of the system to be loss-like, so as to induce non-positive imaginary parts in the effective complex energy eigenvalues. This is easy to see as the poles of the retarded lead Green's function entering Eq. (\ref{eqn:generalHNH}) are constrained to lie in the lower complex half-plane. However, we note that practical experiments on photonic waveguides also work with passive systems only exhibiting loss, which amounts to a constant shift of a PT symmetric spectrum along the imaginary axis \cite{wiemannkremerplotniklumernoltemakrissegevrechtsmanszameit}.   

Exceptional points accompanied by arc-shaped degeneracies have recently been observed in impressive light scattering experiments with photonic crystals in a classical regime \cite{NHarc}. Our present work proposes an experimentally feasible platform that paves the way towards exploring the counterpart of such genuinely NH phenomena in quantum transport and spectroscopy experiments on electronic quantum many-body systems. Quite remarkably, while the arc structures observed in photonic crystals are minuscule as compared to the Brillouin zone of the crystalline structure, here we find a quite sizable splitting of the exceptional degeneracies in reciprocal space, which speaks for the quantitatively robust nature of the predicted phenomena.

Complementary to our present approach, intriguing proposals for NH topological phases in quantum condensed matter systems stemming from strong interactions and disorder have been put forward \cite{koziifu,yoshidapeterskawakmi,Zyuzin2018,papaj,Yoshida,Zyuzin2019,Moors}. However, it is fair to say that the quantitative relevance of those NH effects in real material systems remains to be demonstrated. In contrast, the materials junctions studied in our present work provide a well controlled and experimentally tuneable platform for the realization of NH topological phases. We note that interesting NH physics induced by a lead self-energy has been reported in 1D (proximity induced) superconductor metal junctions in Refs. \cite{MadridMajoranaEP,MajoranaEPPreprint,Pikulin}, where the occurrence of surface states reminiscent of Majorana bound states has been explained in terms of an exceptional point transition that occurs as a function of external parameters characterizing the 0D junction. Moreover, Ref. \onlinecite{Philip} considered a setup similar to ours but focused instead on the gapped anomalous Hall region and found that non-Hermitian effects destroy the quantization of the Hall conductance.    
Here, by contrast, we are concerned with metallic 2D surfaces effectively described by NH models that in a whole parameter range contain exceptional points and Fermi arcs connecting them as a function of the conserved momenta parallel to the junction interface, thus defining stable NH topological phases.  

\acknowledgments
{\it Acknowledgments.---}
We would like to thank Carsten Timm for helpful comments as well as Jorge Cayao for discussions on related topics. E.J.B. is supported by the Swedish Research Council (VR) and the Wallenberg Academy Fellows program of the Knut and Alice Wallenberg Foundation. J.C.B. acknowledges financial support from the German Research Foundation (DFG) through the Collaborative Research Centre SFB 1143.


\begin{thebibliography}{10}
\bibitem{Foot1}
Specifically, we verified that the frequency dependence of the self-energy in the considered parameter regimes is weak enough not to introduce any qualitative changes to our zero frequency approximation.
\end{thebibliography}
\end{document}